\documentstyle[12pt]{article}
\newcommand {\be}       {\begin{equation}}
\newcommand {\ee}    {\end{equation}}
\newcommand {\bea}      {\begin{eqnarray}}
\newcommand {\eea}  {\end{eqnarray}}
              
\newcommand {\bib}[1]   {\bibitem{#1}}

\newcommand {\sqbr}[1]  {\left[ #1 \right] }
\begin{document}
\begin{center} 
\section*{Application of the Coupled Cluster Method
to a Hamiltonian Lattice Field Theory}

 D.Sch\"{u}tte, A. Wichmann and C. Weichmann\\
Institut f\"{u}r Theoretische Kernphysik der Universit\"{a}t Bonn \\
         Nu{\ss}allee 14 -- 16, D-53115 Bonn, Germany \\
\end{center}

\begin{abstract}
The coupled cluster method
has been applied to 
the eigenvalue problem 
lattice Hamiltonian QCD (without quarks)
for SU(2) gauge fields in two space dimensions.

Using a recently presented new formulation
and the truncation prescription of Guo et al.
we were able to compute the ground state and
the lowest $0^+$-glueball mass up to
the sixth order of the coupled cluster expansion.

The results show  evidence for
a ``scaling window'' 
(i.e. good convergence and constance of
dimensionless quantities) around $\beta=4/g^2 \approx 3$.

A  comparison of our results to 
those of other methods is presented. 

\end{abstract}

\subsection*{1 Introduction}
It is the purpose of this paper to present our
first results of an attempt to compute
the spectrum of a lattice gauge field
theory within the Hamiltonian formulation.
Our computational framework is the
coupled cluster method which has been
presented and discussed in detail in Ref.\cite{Schutte}.

We have applied our method to the lattice version of 
the 2+1 dimensional $SU(2)$ Yang-Mills theory which
is the simplest non-trivial non-abelian
lattice gauge field theory.
We consider this model as an important test case
for  controlling our computational scheme.
Although there is no experiment to compare with,
there exist for this model very reliable 
results for the ground state energy\cite{Green}
and for glueball masses\cite{Teper} which allow a 
critical test of the coupled cluster predictions.

\subsection*{2 Computational Scheme}

Within the Hamiltonian formulation, 
 the computational framework
is given by the Kogut Susskind theory
which has been discussed in detail  
in Ref.\cite{Schutte}. In summary, the structure
is the following: 

The Kogut Susskind
wave functions $\Psi(U) = \Psi(U_1,..,U_N)$ 
depend on link variables $U_l$ 
 ($l = 1,..,N$) which are
 elements of the gauge group
$SU(2)$.  $N$ is the number of 
oriented links in a 2-dimensional  lattice
and is related to the chosen
 finite volume. 
Our many-body techniques allow us to
perform the final calculations in
the infinite volume limit $N \rightarrow \infty$.

The idea of the coupled cluster method
is to reformulate the eigenvalue problem
$H \Psi = E \Psi$
as equations for the ground state correlation function
$S$ and for the excitation operators $F$,
 following from the ansatzes
$\Psi_0(U) = \exp{S(U)}$ for the ground state and
$ \Psi(U) = F(U) \exp{S(U)}$
for excited states.

If $H = 1/(2a)(\sum_{lk}g^2 E^2_{lk} -2/g^2V)$ 
is the Kogut Susskind Hamiltonian
(we use the notations $(lk)$ 
for link-colour quantum numbers, ($k = 1,..3$),
 $E_{lk}$ for  the  momentum operators conjugate to $U_l$
and $V  $ for the plaquette term.)
this yields the
non-linear equation
\be
\sum_{lk} (S_{{lk}{lk}} + S_{lk} S_{lk} )
   - {2 \over g^4} V =  {2a \over g^2} E_0
\label{e23}
\ee
for $S$ and  the linear equation
\bea
\sum_{lk} (F_{{lk}{lk}} + 2 S_{lk} F_{lk}) &=&
{2a \over g^2}  (E-E_0) F\;\;
\label{e24}
\eea
for the excitation operators $F$. 
 $f_{lk}$ abbreviates ``link variable derivatives''
of  functions $f(U)$:
\be
f_{lk} = \sqbr{E_{lk},f}\;\;,\;\;
  f_{{lk}{lk}} = \sqbr{E_{lk},\sqbr{E_{lk},f}}\;\;.
\ee
 
This (rigorous) form of the eigenvalue problem guarantees
manifestly the correct volume dependencies
of the ground state energy $E_0$ 
and the excitation energies $E - E_0$ (see Ref.\cite{Schutte}).

Within this paper we will restrict ourselves 
to excitations corresponding to glueballs
defining a trivial representation of the
lattice Euclidean group
($0^+$ glueballs with momentum zero). 
In this case we may write
\be
F =  \Pi_0 F_{\rm int}
\;\;\;\;\;\;\;\;\;\;S = \Pi_0 S_{\rm int}
\label{e27}
\ee
where $ \Pi_0$ is the projection operator on states
with vanishing (lattice) momentum, angular 
momentum  and parity plus.

A calculation for general representations of the
lattice Euclidean group is possible (see\cite{Schutte})
but technically more involved. As a first test for our
method we have chosen  the simplest case.
Our results encourage us to try the generalization and
we are doing this now, but we will defer the presentation
of results to a forthcoming publication.

The essential point of the coupled cluster method
is  to {\em expand} the 
intrinsic functions $S_{\rm int}$ and $F_{\rm int}$
with respect to a 
gauge invariant, linked, standardized  basis $\chi_\alpha$ 
in the form
\be
S_{\rm int}(U) = \sum_\alpha S_\alpha \chi^\alpha(U)
\;\;\;\;\;\;
F_{\rm int}(U) = \sum_\alpha 
F_\alpha \chi^\alpha(U)
\label{e37}
\ee
 and
to define  approximations by  {\em truncations} of this basis.

Introducing the constant function via
$\chi^0 = 1$, 
the ``plaquette
function''  by putting 
$\Pi_0 \chi^1 = 4 V$ and using the strong coupling structure
 $ \sum_{la}  \chi^\alpha_{la la} =
\epsilon_{\alpha} \chi^\alpha$,
the coupled cluster equations
 (\ref{e23}) and (\ref{e24})  become
 equations for the coefficients $S_\alpha$ 
and $F_\alpha$ 
\bea
\epsilon_\alpha S_\alpha + \sum_{\beta}  
C^{\beta}_\alpha(S) S_\beta &=& 
\frac{1}{2 g^4} \delta_{\alpha 1} 
           +\frac{aE_0}{4 N g^2} \delta_{\alpha 0}
\label{e38}
\eea
\bea 
    \epsilon_\alpha F_\alpha + 
  2 \sum_{\beta} 
C^{\beta}_\alpha(S) F_\beta &=& 
           {2a \over g^2}(E-E_0) F_\alpha
 \label{e38a}
\\ \nonumber with\;\;\;\;\;\;\;\;\;
C^{\beta}_\alpha(S)  &=&\sum_\gamma C^{\gamma\beta}_\alpha S_\gamma\;\;.
\eea
Here, the coupled cluster matrix elements
$C^{\gamma\beta}_\alpha$ 
are given by
\bea
C^{\gamma \beta}_\alpha  &=&
 \frac{1}{2}(\epsilon_{\alpha} -
\epsilon_{\beta} - \epsilon_{\gamma})
\sum_{u} c^{\gamma\beta}_{\alpha u} 
\label{e40}
\eea
where the
 numbers $c^{\beta\gamma}_{\alpha u}$
are related to the action $T(u)$ ($u \in G_E$) of the
lattice Euclidean group $G_E$ on 
the basis $\chi^\alpha$  and it's  products by
\be
\sum_{u\in G_E} 
\chi^\beta T(u) \chi^\gamma
= \sum_{\alpha,u}
 c^{\beta\gamma}_{\alpha u} T(u)\chi^\alpha 
\label{e39}
\ee
Hereby, only those cases
have to be considered where the functions $\chi^\beta$ and 
$T(u) \chi^\gamma$ have a common link variable.
Details are given in \cite{Schutte}. 
\\

The basis $\chi^\alpha$ which we use for 
our calculations 
is systematically generated from (\ref{e39})
and given by the following prescriptions.

Introduce first a set of ``generic'' functions
$\Lambda_G^{\delta,k}$
given by  linked,
standardized $\delta$-fold  plaquette products.
They have the structure
\be
\Lambda_G^{\delta,k}=\chi^1 T(u_2(\delta,k))
\chi^1 ...T(u_\delta(\delta,k))\chi^1
\;\; ; \;k=1,..,n_\delta
\label{e53}
\ee
and we define  $n_0 = n_1 = 1$, 
$\Lambda_G^{0,1} = \chi^0$, $\Lambda_G^{1,1} = \chi^1$.

$\delta (=0,1,2...)$ is the ``order'' 
of the function $\Lambda_G^{\delta,k}$.

For our two-dimensional SU(2) case we have 
$n_\delta = 1,1,2,4,12,35,129 $ 
up to sixth order, respectively.
The functions (\ref{e53}) are characterized by simple
loop patterns 
exemplified in \cite{Schutte}.

The essential property of this set of functions is 
that from them one can construct an  orthogonal
(in the limit $\delta \rightarrow \infty$ complete)
basis $\chi^\alpha$ by acting with suitable
Casimir operators of the lattice orthogonal group
on  $\Lambda_G^{\delta,k}$
and diagonalizing 
the corresponding Casimir matrices.
 This basis is
then (iteratively) ``ordered'' by $\delta$ 
because the Casimir operators
do not increase $\delta$, i.e. each element $\chi^\alpha$
has some order $\delta(\alpha)$.

Up to order six, one obtains in this manner subspaces
of the function space of link variables of dimensions
$(1,2,5,15,84,557,4972)$,
respectively. The relation to the simple generic
functions (\ref{e53}), 
which is set up by diagonalizing the
Casimir matrices numerically, allows  then
to compute also the coefficients
$c^{\beta\gamma}_{\alpha u}$ (see \cite{Schutte}).

When constructing  the basis $\chi_\alpha$ in the way
described, one has to take care of possible
linear dependencies between the generated functions.
In previous investigations \cite{Llew,Guo,Luo,Schutte}
this problem was solved by exploiting the
Cayley Hamilton relation between matrices.
We have used in this connection a much simpler procedure: 
If a set  $f_1(U),..,f_n(U)$
contains only $m$ ($m \leq n$) independent
functions, the 
matrix $f_i(U^k) (i,k = 1,..,n) $
has for suitable {\em fixed} variables
$(U^1,..,U^n)$ exactly the rank $m$.
Our experience is that statistically chosen
variable sets $(U^1,..,U^n) \in SU(2)$ are suitable in this
sense. The linear relation between the
functions $f_i$ is then easily constructed 
and the dependent functions can be eliminated.

Having determined in this way all ingredients
of the coupled cluster equations (\ref{e38}) and (\ref{e38a}),
we still have to define a truncation prescription.
Up to now, we have  used in this
connection the proposal of  Guo et al\cite{Guo}
which yields actually  the simplest
set of equations. In this case one puts
in the order $\delta$
\be
c^{\alpha_1,\alpha_2}_{\alpha_3,u} = 0
\;\;\mbox{ for }\;\;\; \delta(\alpha_1) + \delta(\alpha_2) > \delta
\ee

\subsection*{ 3 Results and Discussion}
\subsubsection*{3.1 The Ground State}
It is a special feature of the Hamiltonian 
formulation  that is provides the energy and the wave
function of the vacuum state which
is the ground state of the  Hamiltonian.

Standard lattice Monte Carlo calculations
do not give results for the vacuum energy
density. There exist, however, computations
within the strong coupling expansion\cite{Strong} and
very reliable Green's function Monte Carlo results\cite{Green}.

In Figure 1 we compare our coupled cluster
results up to sixth order to the results of the
other methods. As coupling variable we
 use the standard expression
$\beta = 4/g^2$.
We see that we  have a good quality and
convergence up to $\beta \approx 4$,
but our method breaks down for large $\beta$
where the Green's function Monte Carlo is 
still valid.

An important feature of our results is that the
validity of the coupled cluster method clearly 
goes beyond the range of the strong coupling
expansion which breaks down at $\beta \approx 3$.
(Our figure gives the 18th order of strong
coupling perturbation theory!) We consider
this improvement as a necessary condition for obtaining
continuum limit physics.

We should stress  that the determination
of the vacuum operator $S(U)$ 
- by solving (\ref{e38})  iteratively -
turns out be
be much simpler and faster than that of
the excitation operator $F(U)$.
In this sense we do not see any special
difficulty with the fact that the
Hamiltonian formulation also involves
the determination of the vacuum state.
The coupled cluster formulation deals
with this problem apparently quite effectively.

Note also that assuming a trivial (strong coupling)
 vacuum (i.e. $S(U) = 0$) 
makes no sense within our framework,
this would only yield the completely
unphysical strong coupling limit spectrum.

In this sense our framework seems to be
quite orthogonal to the light front formulation
which is based on the assumption that
the simplicity of the vacuum could be
helpful for the determination of the spectrum.

\subsubsection*{3.2 The Glueball} 

The solution of (\ref{e38a}) with the lowest eigenvalue
has the interpretion of an (approximate) $0^+$ glueball mass.

Within the standard lattice Monte Carlo method,
a  reliable determination of this glueball mass
has been achieved by Teper\cite{Teper}.

A comparison to our Hamiltonian results
involves a rescaling 
of the standard ``Euclidean''  coupling $g_E$
relative to the Hamiltonian coupling $g$ given by
\be
\beta_{E} = \beta + .077 + O(g^2)
\ee
and a rescaling of the Euclidean masses $M_E$ 
relative to  the Hamiltonian masses $M$
(``velocity of light correction'')
\be
M_{E} = (1 + 0.084 g^2
+O(g^4)) M
\ee
(See Ref.\cite{RESC}).

According to Teper's results,
the region of asymptotic scaling starts
at $\beta_{E} \approx 3$
for the string tension and for 
a somewhat larger value of $\beta_E$
for the glueball  masses.

Since our ground state results are
reliable up to $\beta \approx 4$,
one may hope for a scaling window 
for our glueball predictions in this range.

A direct comparison to Teper's results
is difficult in this $\beta$-range 
because the lowest order relations (12) and (13)
are not sufficient for this case.
These corrections are
large in the strong coupling regime itself
because the strong coupling expansions
are completely different in the 
Euclidean and Hamiltonian formalisms;
the Euclidean expansion has a logarithmic
singularity for $\beta = 0$ which
is not present in the Hamiltonian framework.

Therefore, 
the best test for a window displaying continuum physics
is the computation of dimensionless
quantities which should become independent
of $\beta$. Since we have (up to now)
only one glueball at our disposal
(higher $0^+$-glueball masses turn out be be
rather unreliable), we have determined 
within the same coupled cluster approach
the ``radius'' $R$ of
the glueball defined
by
\be
R^2 = \left(\sum_\alpha |F_\alpha|^2|\chi^\alpha|^2 N(\chi^\alpha)\right)
/(\sum_\alpha |F_\alpha|^2 |\chi\alpha|^2)
\label{rad}\ee
where  $N(\chi^\alpha)$ is the number of {\em different}
plaquettes occuring in the corresponding generating
function (\ref{e53}). (The Haar measure norms $|\chi^\alpha|$
were calculated using a Monte Carlo method.)

The result for dimensionless quantity $RM$
is given in Figure 2. There is apparently
evidence for a scaling window around
$\beta \approx 3$.

In Figure 3, we also give our
results for the absolute values of the
glueball masses. Most interesting is
a comparison to the predictions of
the (Hamiltonian) strong coupling
expansion\cite{Strong} and of  corresponding extrapolations
(ELCE-method)\cite{ELCE} which are 
based on the same Kogut Susskind Hamiltonian.

While the strong coupling expansion breaks down
at $\beta \approx 2.1$
(seen in Figure 3 from the 28th order
of this expansion) our results
show a clear convergence (up to order six)
up to $\beta \approx 3$.
Thus we are able to make reliable
calculations in the range  which is beyond, but not
too far beyond the radius of convergence
of the strong coupling expansion.
This is actually the typical range of
the reached scaling window in most
Monte Carlo lattice QCD calculations.

Comparing to the ELCE method, we see
a clear discrepancy already in the
range $\beta \approx 3$.
The ELCE extrapolation  seems
to overestimate the mass values
and it is difficult to understand how this result
will converge to the 
asymptotic glueball
mass, given by Teper as $M(g \rightarrow 0)/2g^2 = 3.2$
with a $1\%$ error.

The coupled cluster method, on the
other hand, appears to give a prediction
that does not contradict to
 this 
continuum limit value, although
a direct comparison to Teper's
result is hampered by the lack
of good knowledge
of the rescaling corrections between
Hamiltonian and Lagrangian formulation
for $\beta \approx 3$.

Summarizing we conclude that the
coupled cluster method appears to
be able to produce predictions
within Hamiltonian lattice field
theories in a range of the couplings
where asymptotic scaling is already valid.
Hereby, the simultaneous computation
of the vacuum state poses no special
difficulty, 
it may be reliably determined in the  region 
where observables show a scaling behaviour
(and beyond,
i.e. up to $\beta \approx 4$ in our case).
Also the numerical effort is much less 
for the vacuum than
for the diagonalization problem of the
glueball masses.

Clearly, a more decisive test of our
computational framework will
be our predictions for glueball
masses with more general (lattice)
angular momentum quantum numbers
and the comparison of the emerging
mass ratios to those of Teper.
Calculations in this direction are
now being undertaken.

We would like to thank Bernhard Metsch for
many valuable discussions.
D. S. also acknowledges a useful exchange of ideas
with Chris Hamer, Helmut Kr\"oger and Xiang Luo.

\clearpage

\section*{Figure Captions}

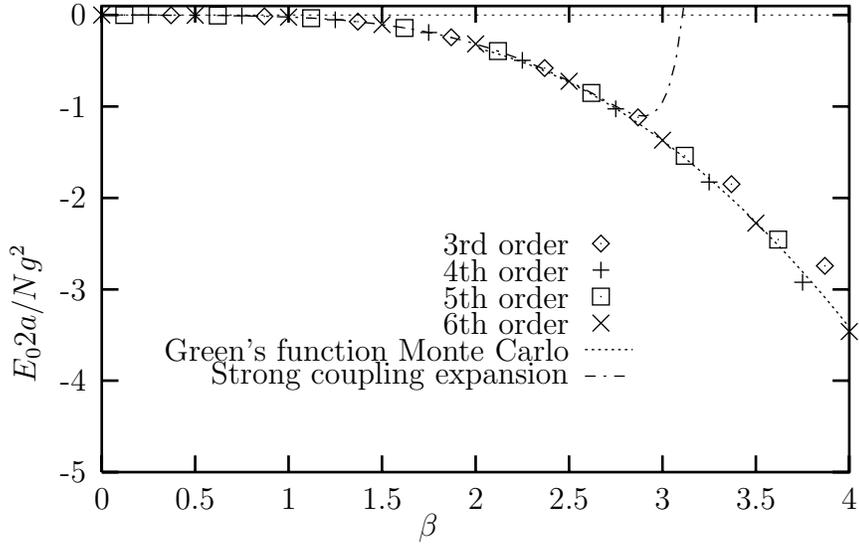
\begin{figure}[htbp]
  \caption{Coupled cluster predictions of this work
for the ground state energy density in 3rd, 4th, 5th and 6th order, 
compared with the 18th order of the 
strong coupling expansion \protect\cite{Strong} 
and the results of the Green's function 
Monte Carlo method  \protect\cite{Green}.
The upper part magnifies the low $\beta$ range of the results
to show the 
quality of the convergence.}
\end{figure}

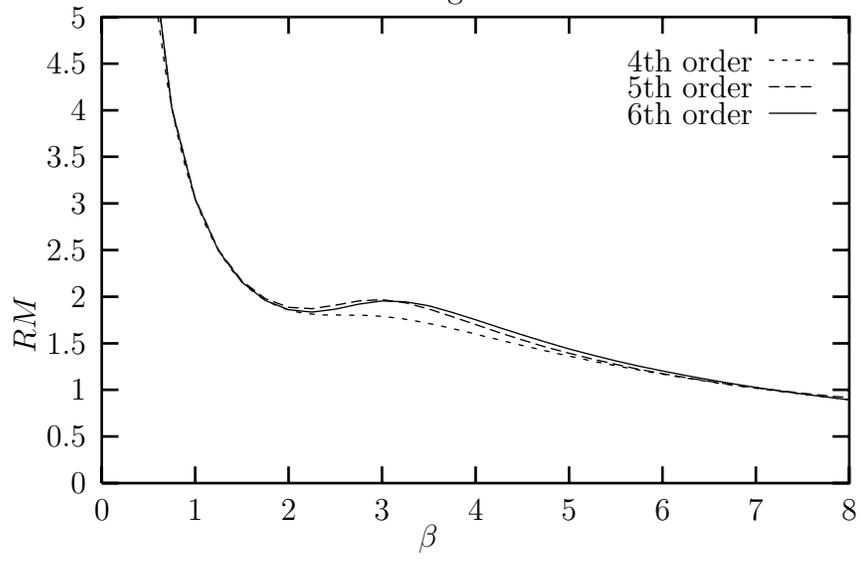
\begin{figure}[htbp]
    \caption{The dimensionless product of glueball mass and radius
$MR$ in 4th, 5th and 6th order.}
\end{figure}

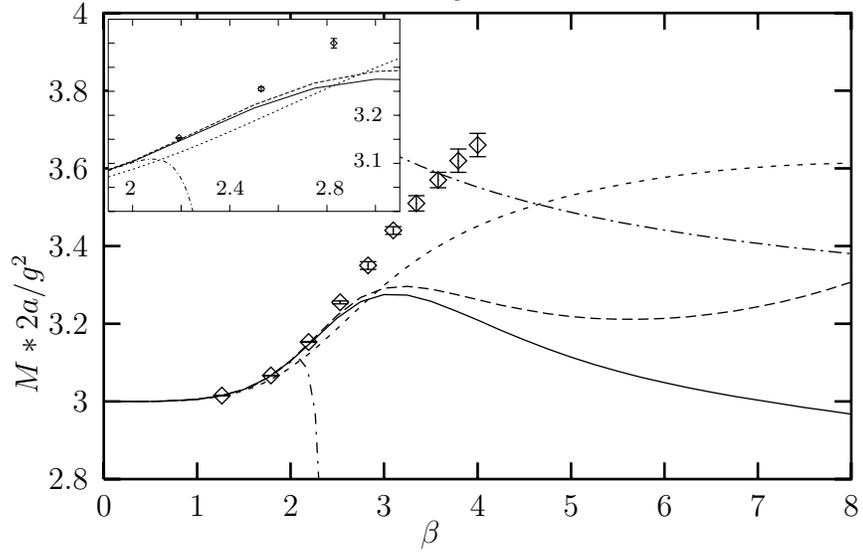
\begin{figure}[htbp]
    \caption{Mass $M$ of the $0^+$ glueball in 4th 
(
\hspace{18bp}),
      5th (
\hspace{18bp}) and 6th (
\hspace{18bp}) order in comparison
      with the 28th order of
the strong coupling expansion \protect\cite{Strong} (lower
      dot-dashed line) and an extrapolation using
 the ELCE method  \protect\cite{Strong} (diamonds with error bars).
 The upper dot-dashed line gives 
     the Monte Carlo results of Teper  \protect\cite{Teper}  
using a lowest order Euclidean-Hamiltonian rescaling 
given by eqs. (12) and (13).
The Monte Carlo errors (1-3\%) are small compared to the
uncertainty of these rescaling corrections
(see text) and are therefore not
indicated in the figure. 
      The insert shows a magnification of the crucial part between $\beta=2$ and $\beta=3$.}
\end{figure}

\clearpage

\setlength{\unitlength}{0.1bp}
\begin{picture}(3600,2160)(0,0)
\put(2361,613){\makebox(0,0)[r]{Strong coupling expansion}}
\put(2361,713){\makebox(0,0)[r]{Green's function Monte Carlo}}
\put(2361,813){\makebox(0,0)[r]{6th order}}
\put(2361,913){\makebox(0,0)[r]{5th order}}
\put(2361,1013){\makebox(0,0)[r]{4th order}}
\put(2361,1113){\makebox(0,0)[r]{3rd order}}
\put(2008,2109){\makebox(0,0){Figure 1}}
\put(1830,40){\makebox(0,0){$\beta$}}
\put(350,850){%
\makebox(0,0)[b]{\shortstack{$E_02a/Ng^2$}}%
}
\put(3417,151){\makebox(0,0){4}}
\put(3065,151){\makebox(0,0){3.5}}
\put(2713,151){\makebox(0,0){3}}
\put(2361,151){\makebox(0,0){2.5}}
\put(2009,151){\makebox(0,0){2}}
\put(1656,151){\makebox(0,0){1.5}}
\put(1304,151){\makebox(0,0){1}}
\put(952,151){\makebox(0,0){0.5}}
\put(600,151){\makebox(0,0){0}}
\put(540,1975){\makebox(0,0)[r]{0}}
\put(540,1630){\makebox(0,0)[r]{-1}}
\put(540,1285){\makebox(0,0)[r]{-2}}
\put(540,940){\makebox(0,0)[r]{-3}}
\put(540,596){\makebox(0,0)[r]{-4}}
\put(540,251){\makebox(0,0)[r]{-5}}
\end{picture}

\vspace{2cm}
\setlength{\unitlength}{0.1bp}
\begin{picture}(3600,2160)(0,0)
\put(2290,646){\makebox(0,0)[r]{Green's function Monte Carlo}}
\put(2290,746){\makebox(0,0)[r]{6th order}}
\put(2290,846){\makebox(0,0)[r]{5th order}}
\put(2290,946){\makebox(0,0)[r]{4th order}}
\put(2290,1046){\makebox(0,0)[r]{3rd order}}
\put(2008,40){\makebox(0,0){$\beta$}}
\put(350,850){%
\makebox(0,0)[b]{\shortstack{$E_02a/Ng^2$}}%
}
\put(3417,151){\makebox(0,0){10}}
\put(2854,151){\makebox(0,0){8}}
\put(2290,151){\makebox(0,0){6}}
\put(1727,151){\makebox(0,0){4}}
\put(1163,151){\makebox(0,0){2}}
\put(600,151){\makebox(0,0){0}}
\put(540,2106){\makebox(0,0)[r]{0}}
\put(540,1841){\makebox(0,0)[r]{-10}}
\put(540,1576){\makebox(0,0)[r]{-20}}
\put(540,1311){\makebox(0,0)[r]{-30}}
\put(540,1046){\makebox(0,0)[r]{-40}}
\put(540,781){\makebox(0,0)[r]{-50}}
\put(540,516){\makebox(0,0)[r]{-60}}
\put(540,251){\makebox(0,0)[r]{-70}}
\end{picture}

\clearpage
\setlength{\unitlength}{0.1bp}
\begin{picture}(3600,2160)(0,0)
\put(3054,1646){\makebox(0,0)[r]{6th order}}
\put(3054,1746){\makebox(0,0)[r]{5th order}}
\put(3054,1846){\makebox(0,0)[r]{4th order}}
\put(2008,2109){\makebox(0,0){Figure 2}}
\put(1830,40){\makebox(0,0){$\beta$}}
\put(350,850){%
\makebox(0,0)[b]{\shortstack{$RM$}}%
}
\put(3417,151){\makebox(0,0){8}}
\put(3065,151){\makebox(0,0){7}}
\put(2713,151){\makebox(0,0){6}}
\put(2361,151){\makebox(0,0){5}}
\put(2009,151){\makebox(0,0){4}}
\put(1656,151){\makebox(0,0){3}}
\put(1304,151){\makebox(0,0){2}}
\put(952,151){\makebox(0,0){1}}
\put(600,151){\makebox(0,0){0}}
\put(540,2009){\makebox(0,0)[r]{5}}
\put(540,1833){\makebox(0,0)[r]{4.5}}
\put(540,1657){\makebox(0,0)[r]{4}}
\put(540,1482){\makebox(0,0)[r]{3.5}}
\put(540,1306){\makebox(0,0)[r]{3}}
\put(540,1130){\makebox(0,0)[r]{2.5}}
\put(540,954){\makebox(0,0)[r]{2}}
\put(540,778){\makebox(0,0)[r]{1.5}}
\put(540,603){\makebox(0,0)[r]{1}}
\put(540,427){\makebox(0,0)[r]{0.5}}
\put(540,251){\makebox(0,0)[r]{0}}
\end{picture}

\clearpage
\setlength{\unitlength}{0.1bp}
\begin{picture}(3600,2160)(0,0)
\put(2008,2109){\makebox(0,0){Figure 3}}
\put(1830,40){\makebox(0,0){$\beta$}}
\put(350,837){%
\makebox(0,0)[b]{\shortstack{$M*2a/g^2$}}%
}
\put(3417,151){\makebox(0,0){8}}
\put(3065,151){\makebox(0,0){7}}
\put(2713,151){\makebox(0,0){6}}
\put(2361,151){\makebox(0,0){5}}
\put(2009,151){\makebox(0,0){4}}
\put(1656,151){\makebox(0,0){3}}
\put(1304,151){\makebox(0,0){2}}
\put(952,151){\makebox(0,0){1}}
\put(600,151){\makebox(0,0){0}}
\put(540,2009){\makebox(0,0)[r]{4}}
\put(540,1716){\makebox(0,0)[r]{3.8}}
\put(540,1423){\makebox(0,0)[r]{3.6}}
\put(540,1130){\makebox(0,0)[r]{3.4}}
\put(540,837){\makebox(0,0)[r]{3.2}}
\put(540,544){\makebox(0,0)[r]{3}}
\put(540,251){\makebox(0,0)[r]{2.8}}
\scriptsize
\put(1452,1350){\makebox(0,0){2.8}}
\put(1076,1350){\makebox(0,0){2.4}}
\put(700,1350){\makebox(0,0){2}}
\put(1650,1628){\makebox(0,0)[r]{3.2}}
\put(1650,1454){\makebox(0,0)[r]{3.1}}
\end{picture}

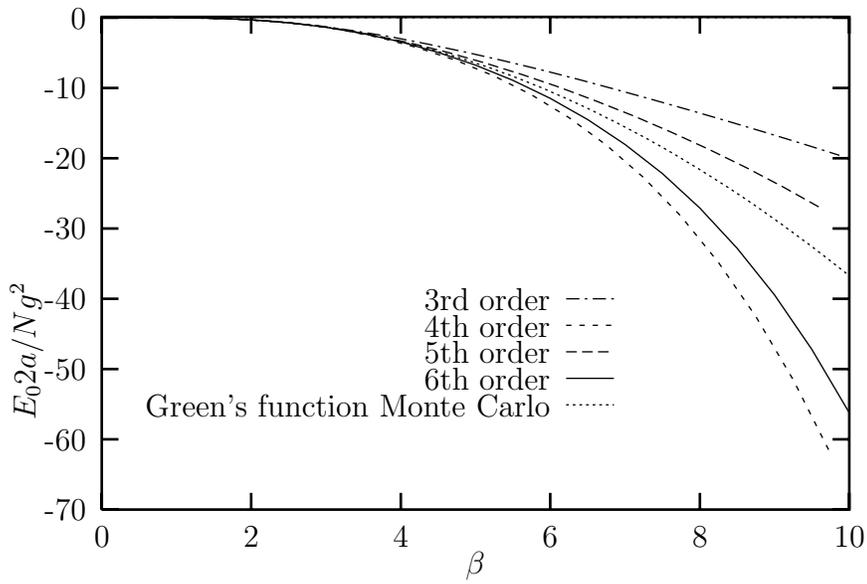
\end{document}